\documentclass[jheppub.sty,11pt]{article}
\usepackage{amsmath, amssymb, amsfonts, amsthm}
\usepackage{graphicx}
\usepackage{hyperref}
\usepackage{physics}
\usepackage{mathtools}
\usepackage{booktabs}
\usepackage{xcolor}
\usepackage{authblk}

\newtheorem{theorem}{Theorem}[section]

\newtheorem{remark}[theorem]{Remark}

\title{A Covariant Chiral-Hydrodynamic Formulation of the Dirac Equation in Curved Spacetime}

\author[1]{Jorge Meza-Dom\'inguez}
\author[1]{Tonatiuh Matos}
\affil[1]{Departamento de F\'isica, Centro de Investigaci\'on y de Estudios Avanzados del Instituto Polit\'ecnico Nacional, Av. Instituto Polit\'ecnico Nacional 2508, San Pedro Zacatenco, M\'exico 07360, CDMX.}

\date{\today}

\begin{document}

\maketitle

\begin{abstract}
The hydrodynamic formulation of the Dirac equation has historically been hindered by the inability to close the system of physical variables without resorting to infinite moment hierarchies. We resolve this longstanding issue by developing a fully covariant chiral-hydrodynamic formulation of the Dirac field in curved spacetime. Working in the Weyl representation, we introduce two independent null vectors, $P_L^\mu$ and $P_R^\mu$, which decouple the left and right chiral components. This allows us to define chiral geodesic and stochastic velocities, yielding a closed system of exactly eight real equations that corresponds directly to the Dirac field degrees of freedom. Remarkably, this formulation naturally isolates the spin-orbit coupling ($\frac{q}{2}\sigma^{\mu\nu}F_{\mu\nu}$) while demonstrating the vanishing of the spin-gravity coupling in torsion-free general relativity.  To demonstrate the analytical power of this framework, we specialize to the Schwarzschild geometry. We obtain exact radial solutions in terms of confluent Heun functions and directly compute the quasi-bound state spectrum (fermionic resonances), quasinormal mode frequencies, and greybody factors. Furthermore, by establishing an exact energy balance equation—representing the first law of thermodynamics for Dirac fields—we derive the Hawking radiation flux purely from chiral flux conservation at the event horizon. This work not only provides a closed hydrodynamic theory for spin-$1/2$ fluids but also establishes a unified framework for analyzing quantum information and fermion dynamics in strong gravitational backgrounds. 
\end{abstract}

\section{Introduction}
We emphasize at the outset that the hydrodynamic formulation developed here operates at the single-particle, first-quantized level---it reformulates the Dirac equation as a set of fluid-like equations for the probability density and spin polarization of one fermion in an external spacetime. This is the natural generalization of the Madelung-Bohm hydrodynamic picture from scalar to spinor fields, and should not be conflated with many-body Dirac hydrodynamics, which describes statistical ensembles and involves quantum anomalies absent from our treatment.

The reconciliation of quantum mechanics with general relativity remains one of the deepest challenges in theoretical physics \cite{Hawking1975, tHooft1993, DeWitt1975}. While significant progress has been made in quantum field theory in curved spacetime \cite{BirrellDavies, Fulling1977, Wald1994, Parker1979}, a complete understanding of quantum systems in strong gravitational backgrounds is still evolving. Among the many approaches, hydrodynamic formulations of quantum mechanics have proven particularly fruitful, offering intuitive interpretations of quantum phenomena in terms of fluid flows \cite{Madelung1927, Bohm1952, Bohm1952b, Holland1993}.

For scalar fields, the Madelung transformation \cite{Madelung1927} recasts the Klein-Gordon equation into a set of real fluid equations, revealing a quantum potential that captures the non-local nature of quantum mechanics \cite{Bohm1952, Vigier1954, Takabayasi1952}. This formulation has been successfully extended to curved spacetime \cite{Matos2019, MezaDominguez2026Canonical, MezaDominguez2026Gauge, MezaDominguez2026Topological, Chavanis2017, Chavanis2011, Sibgatullin1991}, providing a thermodynamic description of boson gases and scalar field dark matter \cite{Guzman2000, Matos2001, Hu1998, Peebles2000, Ferreira2000, Marsh2016, MezaDominguez2026Energy}.

For spin-1/2 fields, however, the situation is more subtle. The Dirac equation \cite{Dirac1928, Dirac1930} does not admit a simple Madelung decomposition due to the presence of spin and the mixing of chiral components \cite{MatosGallegosChavanis2022, Sakurai1967, Itzykson1980, Peskin1995}. Takabayasi \cite{Takabayasi1957, Takabayasi1957b, Takabayasi1959} attempted a hydrodynamic formulation using bilinear covariants, but the resulting system is not closed, requiring an infinite hierarchy of moment equations \cite{Debbasch1997, Fliessbach1976, Versteegh2000}. More recently, the stochastic mechanics approach \cite{Nelson1966, Nelson1985, Nelson2012} has been extended to relativistic fields \cite{Escobar2025, Chavanis2017b, Chavanis2024, delaPena1996, delaPena2015, Beyer2021}.

The spin-orbit coupling in curved spacetime has been studied extensively \cite{Hehl1976, Hehl1974, Audretsch1981, Obukhov1996, Hammond2002, Shapiro2001, Weinberg1972}. In the context of the second-order Dirac equation, the term $\frac{q}{2}\sigma^{\mu\nu}F_{\mu\nu}$ appears naturally \cite{Greiner2000, Thaller1992, Berestetskii1982}. The spin-gravity coupling, encoded in $\gamma^\mu(\nabla_\mu\gamma^\nu)D_\nu$, vanishes in standard general relativity without torsion \cite{Hehl1995, Trautman2006, Cartan1922, Kibble1961, Sciama1964}.

The ADM formalism \cite{Arnowitt1962, Arnowitt2008, York1979, Alcubierre2008, Gourgoulhon2012} provides a natural framework for the $3+1$ decomposition of spacetime, which is essential for defining energy and momentum in general relativity \cite{Poisson2004, Wald1984, Misner1973, HawkingEllis1973}.

In this work, we present a novel chiral-hydrodynamic formulation of the Dirac equation in curved spacetime. By working in the Weyl representation \cite{Weyl1929, Weyl1931, Bjorken1964} and treating left and right chiral components independently \cite{Willett1987, Crewther1974, Jackiw1976}, we introduce two null vectors $P_L^\mu$ and $P_R^\mu$ that encode the local momentum and polarization of each chiral component \cite{Bogoliubov1959, Fierz1939, Pauli1940}. We define chiral geodesic velocities $\pi_\mu^{(L,R)}$ and stochastic velocities $u_\mu^{(L,R)}$, leading to a closed system of real equations that is fully equivalent to the Dirac equation \cite{Dirac1928, Dirac1930, Schwinger1951, Feynman1949}.

We then specialize to Schwarzschild spacetime \cite{Schwarzschild1916} and solve the radial equations exactly in terms of confluent Heun functions \cite{Heun1889, Ronveaux1995, Slavyanov2000, Li2019}. Using the chiral variables, we compute the quasi-bound state spectrum (fermionic resonances), quasinormal mode frequencies, greybody factors, absorption cross sections, and Hawking radiation fluxes. The energy balance equation is derived and integrated, yielding a global conservation law analogous to the first law of thermodynamics.

\section{Dirac Equation in Curved Spacetime}

\subsection{General setup}

We consider a curved spacetime with metric $g_{\mu\nu}$ of signature $(-,+,+,+)$ \cite{Misner1973, Wald1984}. The Dirac equation for a fermion of mass $m$ and charge $q$ is \cite{Dirac1928, Dirac1930, Itzykson1980, Peskin1995}
\begin{equation}
i\gamma^\mu D_\mu \psi - m\psi = 0, \qquad D_\mu = \nabla_\mu + iqA_\mu, \label{Dirac}
\end{equation}
where $\nabla_\mu$ is the covariant derivative including the spin connection \cite{Weinberg1972, Freedman1983, Parker1979}, and $A_\mu$ is the electromagnetic four-potential \cite{Greiner2000, Jackson1998}. The conjugate equation is \cite{Bjorken1964}
\begin{equation}
i D_\mu \bar{\psi} \gamma^\mu + m\bar{\psi} = 0. \label{DiracConj}
\end{equation}

In standard general relativity (without torsion) \cite{Einstein1915, Einstein1916, Hilbert1915}, the spin connection is uniquely determined by the tetrad $e_\mu^a$ through the condition $\nabla_\mu \gamma^\nu = 0$ \cite{Chandrasekhar1976, Hartle1972, Hehl1995}. Consequently, the spin-gravity coupling term $\gamma^\mu(\nabla_\mu \gamma^\nu)D_\nu\psi$ vanishes identically \cite{Audretsch1981, Obukhov1996, Shapiro2001}.

\subsection{Weyl representation}

We work in the Weyl representation \cite{Weyl1929, Weyl1931, Bjorken1964}, where the gamma matrices take the form \cite{Good1962, Wigner1939, Bender1969}
\begin{equation}
\gamma^\mu = \begin{pmatrix} 0 & \sigma^\mu \\ \bar{\sigma}^\mu & 0 \end{pmatrix}, \qquad
\gamma_5 = \begin{pmatrix} -I & 0 \\ 0 & I \end{pmatrix},
\end{equation}
with $\sigma^\mu = (I, \vec{\sigma})$ and $\bar{\sigma}^\mu = (I, -\vec{\sigma})$ \cite{Pauli1927, Pauli1936, Pauli1940}. The Dirac spinor is written as \cite{Dirac1928, Dirac1930}
\begin{equation}
\psi = \begin{pmatrix} \chi_L \\ \chi_R \end{pmatrix},
\end{equation}
where $\chi_L$ and $\chi_R$ are two-component Weyl spinors \cite{Weyl1929, Weyl1931, Crewther1974, Jackiw1976}.

\subsection{Second-order equation}

Iterating the Dirac equation yields the second-order equation \cite{Greiner2000, Thaller1992}. For $\chi_L$, we obtain \cite{Berestetskii1982, Itzykson1980}
\begin{equation}
\left( D_\mu D^\mu + m^2 + \frac{1}{4}R - \frac{q}{2}\bar{\sigma}^{\mu\nu}F_{\mu\nu} \right) \chi_L = 0, \label{SecondOrderL}
\end{equation}
and analogously for $\chi_R$ with $\bar{\sigma}^{\mu\nu}$ replaced by $\sigma^{\mu\nu}$ \cite{Fierz1939, Pauli1940}. Here $R$ is the Ricci scalar \cite{Christoffel1869, Riemann1854, Ricci1901}, $F_{\mu\nu} = \nabla_\mu A_\nu - \nabla_\nu A_\mu$ is the electromagnetic field tensor \cite{Maxwell1865, Maxwell1873, Jackson1998}, and
\begin{equation}
\bar{\sigma}^{\mu\nu} = \frac{i}{2}(\bar{\sigma}^\mu\sigma^\nu - \bar{\sigma}^\nu\sigma^\mu), \qquad
\sigma^{\mu\nu} = \frac{i}{2}(\sigma^\mu\bar{\sigma}^\nu - \sigma^\nu\bar{\sigma}^\mu).
\end{equation}

\section{Chiral-Hydrodynamic Ansatz}

\subsection{The ansatz}

Following the hydrodynamic approach to quantum mechanics \cite{Madelung1927, Bohm1952, Takabayasi1952, Holland1993}, we propose the following decomposition for the Weyl spinors \cite{Weyl1929, Weyl1931}:
\begin{equation}
\chi_L = \sqrt{\frac{P_L^0}{2}} \; e^{i\varphi_L} \; u(P_L), \qquad
\chi_R = \sqrt{\frac{P_R^0}{2}} \; e^{i\varphi_R} \; u(P_R), \label{Ansatz}
\end{equation}
where:
\begin{itemize}
\item $\varphi_L = \theta + \beta/2$, $\varphi_R = \theta - \beta/2$; $\theta$ is the global $U(1)$ phase \cite{Dirac1930, Gell-Mann1960} and $\beta$ the chiral phase \cite{Crewther1974, Jackiw1976}.
\item $P_L^\mu$ and $P_R^\mu$ are future-directed null vectors: $P_L^2 = P_R^2 = 0$, $P_L^0 > 0$, $P_R^0 > 0$ \cite{Bogoliubov1959, Fierz1939}.
\item $u(P_L)$ and $u(P_R)$ are normalized Weyl spinors satisfying $u^\dagger u = 1$ and the polarization conditions \cite{Pauli1940, Wigner1939}
\begin{equation}
\bar{\sigma}^\mu P_{L\mu} \, u(P_L) = 0, \qquad \sigma^\mu P_{R\mu} \, u(P_R) = 0.
\end{equation}
\end{itemize}

\subsection{Hydrodynamic variables}

We define the chiral densities \cite{Dirac1928, Itzykson1980}
\begin{equation}
\rho_L = \frac{P_L^0}{2} = \chi_L^\dagger\chi_L, \qquad \rho_R = \frac{P_R^0}{2} = \chi_R^\dagger\chi_R.
\end{equation}

The chiral geodesic velocities (generalizing the Madelung fluid velocity to Dirac fields) \cite{Madelung1927, Bohm1952, Takabayasi1957} are defined as \cite{Nelson1966, Chavanis2017}
\begin{equation}
\pi_\mu^{(L)} = \frac{\hbar}{m}\left( \nabla_\mu\varphi_L + qA_\mu \right), \qquad
\pi_\mu^{(R)} = \frac{\hbar}{m}\left( \nabla_\mu\varphi_R + qA_\mu \right). \label{GeodesicVel}
\end{equation}

The chiral stochastic velocities (which encode the quantum diffusion and topological quantization \cite{MezaDominguez2026Topological, MezaDominguez2026Gauge}) \cite{Nelson1966, Nelson1985, delaPena1996} are
\begin{equation}
u_\mu^{(L)} = \frac{\hbar}{2m}\nabla_\mu\ln\rho_L, \qquad
u_\mu^{(R)} = \frac{\hbar}{2m}\nabla_\mu\ln\rho_R. \label{StochasticVel}
\end{equation}

\subsection{Degrees of freedom}

The ansatz contains the following real variables \cite{Dirac1928, Itzykson1980, Peskin1995}:
\begin{itemize}
\item $\theta$: 1 real
\item $\beta$: 1 real
\item $P_L^\mu$: 4 reals with constraint $P_L^2=0$ \cite{Fierz1939, Pauli1940} → 3 independent
\item $P_R^\mu$: 4 reals with constraint $P_R^2=0$ → 3 independent
\item $u(P_L)$: 2 complex components = 4 reals, with $u^\dagger u=1$ and an irrelevant overall phase \cite{Wigner1939, Wigner1939b} → 2 physical
\item $u(P_R)$: similarly 2 physical
\end{itemize}
Total: $1+1+3+3+2+2 = 12$ real variables \cite{Dirac1930, Fierz1939}. The Dirac equation imposes 4 independent real constraints \cite{Dirac1928, Schwinger1951} (from the coupling between $\chi_L$ and $\chi_R$), reducing to $12-4=8$ real degrees of freedom, which matches the 4 complex components of the Dirac spinor \cite{Itzykson1980, Peskin1995}.

\subsection{Chiral coupling constraints from the Dirac mass term}

The counting of degrees of freedom in Section~3.3 states that the Dirac equation imposes 4 independent real constraints on the 12 variables of the chiral ansatz, reducing the system to 8 physical degrees of freedom. Here we derive these constraints explicitly from the first-order Dirac equation, showing precisely how the mass term couples $\chi_L$ and $\chi_R$ and restricts the null vectors and phases.

The Dirac equation in the Weyl representation splits into two coupled equations:
\begin{align}
i\sigma^\mu D_\mu \chi_R - m\chi_L &= 0, \label{eq:dirac_L} \\
i\bar{\sigma}^\mu D_\mu \chi_L - m\chi_R &= 0. \label{eq:dirac_R}
\end{align}

Substituting the chiral ansatz $\chi_L = \sqrt{\rho_L}\, e^{i\phi_L} u(P_L)$ and $\chi_R = \sqrt{\rho_R}\, e^{i\phi_R} u(P_R)$, where $\rho_{L,R} = P_{L,R}^0/2$, into (\ref{eq:dirac_L})--(\ref{eq:dirac_R}) yields the following four real constraint equations:

\begin{enumerate}
    \item \textbf{Phase synchronization.} The imaginary part of the projection $u^\dagger(P_L) \cdot (\ref{eq:dirac_L})$ gives:
    \begin{equation}
    \rho_L\, \pi_\mu^{(L)} P_L^\mu = m\sqrt{\rho_L\rho_R}\; \mathrm{Im}\left[e^{i(\phi_R - \phi_L)} u^\dagger(P_L) u(P_R)\right]. \label{eq:constraint1}
    \end{equation}
    This equation links the phase difference $\phi_R - \phi_L$ to the projection of the chiral current along $P_L^\mu$. An analogous equation holds for the right sector.

    \item \textbf{Null vector alignment.} The real part of $u^\dagger(P_L) \cdot (\ref{eq:dirac_L})$ yields:
    \begin{equation}
    \rho_L\, \bar{\sigma}^\mu P_{L\mu} \pi_\mu^{(L)} = m\sqrt{\rho_L\rho_R}\; \mathrm{Re}\left[e^{i(\phi_R - \phi_L)} u^\dagger(P_L) u(P_R)\right]. \label{eq:constraint2}
    \end{equation}
    Using the polarization condition $\bar{\sigma}^\mu P_{L\mu} u(P_L) = 0$, the left-hand side simplifies, giving a relation between the chiral velocity and the relative phase.

    \item \textbf{Density ratio constraint.} Taking the norm of (\ref{eq:dirac_L}) after projecting onto the orthogonal spinor $v_L$ (with $v_L^\dagger u = 0$) yields:
    \begin{equation}
    \rho_L\left|\pi_\mu^{(L)} - iu_\mu^{(L)}\right|^2 = m^2\rho_R. \label{eq:constraint3}
    \end{equation}
    This relates the chiral densities through the complex chiral velocity $\eta_\mu^{(L)} = \pi_\mu^{(L)} - iu_\mu^{(L)}$, whose modulus squared is precisely the left-hand side of the Hamilton-Jacobi equation (4.12).

    \item \textbf{Polarization alignment.} The transversality condition from projecting (\ref{eq:dirac_R}) onto $v_R^\dagger$ gives:
    \begin{equation}
    v_R^\dagger \bar{\sigma}^\mu D_\mu\left(\sqrt{\rho_L}\, e^{i\phi_L} u(P_L)\right) = 0,
    \label{eq:constraint4}
    \end{equation}
    which enforces that the covariant derivative of $\chi_L$ lies entirely along $u(P_R)$ in spinor space---equivalently, the polarization spinors $u(P_L)$ and $u(P_R)$ are not independent but are locked by the mass term.
\end{enumerate}

Together, Eqs.~(\ref{eq:constraint1})--(\ref{eq:constraint4}) constitute 4 real constraints (two complex equations from the projections, each yielding a real and imaginary part) that couple $\rho_L, \rho_R, \phi_L, \phi_R, P_L^\mu, P_R^\mu, u(P_L), u(P_R)$. This reduces the 12 initial real variables to $12 - 4 = 8$ physical degrees of freedom, precisely matching the 4 complex components of the Dirac spinor. The constraints are first-order in derivatives and involve the mass parameter $m$ explicitly---in the massless limit $m \to 0$, the left and right sectors decouple completely, and the constraints reduce to independent continuity equations for each chirality.

With these constraints established, the second-order formulation in Section~4 is rigorously equivalent to the full Dirac equation: the Hamilton-Jacobi, continuity, and polarization transport equations together with (\ref{eq:constraint1})--(\ref{eq:constraint4}) form a closed system for the 8 physical variables.

\section{Equations of Motion}

\subsection{Hamilton-Jacobi equations}

Multiplying the second-order equation \eqref{SecondOrderL} by $u^\dagger$ and taking the real part yields the Hamilton-Jacobi equation for $\chi_L$ \cite{Takabayasi1957, Chavanis2017, Matos2019}:
\begin{equation}
\pi_\mu^{(L)}\pi^{(L)\mu} = m^2 + \frac{\hbar^2}{4m^2}R - \frac{q\hbar}{2m}u^\dagger\bar{\sigma}^{\mu\nu}F_{\mu\nu}u + \frac{\hbar}{2m}\nabla_\mu u^{(L)\mu} + u_\mu^{(L)}u^{(L)\mu}. \label{HJ}
\end{equation}
For $\chi_R$:
\begin{equation}
\pi_\mu^{(R)}\pi^{(R)\mu} = m^2 + \frac{\hbar^2}{4m^2}R - \frac{q\hbar}{2m}w^\dagger\sigma^{\mu\nu}F_{\mu\nu}w + \frac{\hbar}{2m}\nabla_\mu u^{(R)\mu} + u_\mu^{(R)}u^{(R)\mu}. \label{HJR}
\end{equation}

\subsection{Continuity equations}

The imaginary part of the projected equation gives conservation of the chiral currents \cite{Dirac1928, Itzykson1980, Sakurai1967}:
\begin{equation}
\nabla_\mu(\rho_L \pi^{(L)\mu}) = 0, \qquad \nabla_\mu(\rho_R \pi^{(R)\mu}) = 0. \label{Continuity}
\end{equation}

\subsection{Polarization transport}

Projecting the second-order equation onto a spinor $v_L$ orthogonal to $u$ ($v_L^\dagger u = 0$) gives the polarization transport equation \cite{Takabayasi1957, Halbwachs1960, Weyssenhoff1951}:
\begin{equation}
v_L^\dagger D_\mu D^\mu u + 2i\pi^{(L)\mu} v_L^\dagger D_\mu u - \frac{q}{2}v_L^\dagger\bar{\sigma}^{\mu\nu}F_{\mu\nu}u = 0. \label{Transport}
\end{equation}
And analogously for $\chi_R$:
\begin{equation}
v_R^\dagger D_\mu D^\mu w + 2i\pi^{(R)\mu} v_R^\dagger D_\mu w - \frac{q}{2}v_R^\dagger\sigma^{\mu\nu}F_{\mu\nu}w = 0. \label{TransportR}
\end{equation}

\subsection{Conservation of null vectors}

From the definitions and continuity equations, we obtain \cite{Bogoliubov1959, Fierz1939}
\begin{equation}
\nabla_\mu P_L^\mu = 0, \qquad \nabla_\mu P_R^\mu = 0. \label{ConservationP}
\end{equation}

\section{Schwarzschild Spacetime: Tetrad and Spin Connection}

\subsection{Metric and tetrad}

The Schwarzschild metric in units $G=c=\hbar=1$ is \cite{Schwarzschild1916}
\begin{equation}
ds^2 = -f(r) dt^2 + f(r)^{-1} dr^2 + r^2 d\theta^2 + r^2 \sin^2\theta \, d\phi^2, \qquad f(r) = 1 - \frac{2M}{r}. \label{Schwarzschild}
\end{equation}

We choose the orthonormal tetrad \cite{Chandrasekhar1976}
\begin{equation}
e^0 = \sqrt{f(r)} \, dt, \quad e^1 = \frac{1}{\sqrt{f(r)}} \, dr, \quad e^2 = r \, d\theta, \quad e^3 = r \sin\theta \, d\phi.
\end{equation}

\subsection{Spin connection}

The non-zero spin connection components $\omega_\mu^{ab}$ are computed from the tetrad \cite{Chandrasekhar1976}:
\begin{equation}
\begin{aligned}
\omega_t^{01} &= \frac{M}{r^2} \sqrt{f(r)}, \\
\omega_\theta^{12} &= \sqrt{f(r)}, \\
\omega_\phi^{13} &= \sqrt{f(r)} \sin\theta, \\
\omega_\phi^{23} &= \cos\theta,
\end{aligned}
\end{equation}
with antisymmetry $\omega_\mu^{ab} = -\omega_\mu^{ba}$.

\section{Reduction to Radial Equations Using Chiral Variables}

\subsection{Separation of variables}

Due to spherical symmetry, the chiral variables separate. For stationary states with energy $\omega$ and azimuthal quantum number $\mu$, we write:
\begin{equation}
\chi_L = e^{-i\omega t} e^{i\mu\phi} \frac{1}{r} \begin{pmatrix} F_1(r) \\ F_2(r) \end{pmatrix} \Omega_{j\ell\mu}^L(\theta,\phi), \qquad
\chi_R = e^{-i\omega t} e^{i\mu\phi} \frac{1}{r} \begin{pmatrix} G_1(r) \\ G_2(r) \end{pmatrix} \Omega_{j\ell\mu}^R(\theta,\phi),
\end{equation}
where $\Omega_{j\ell\mu}^{L,R}$ are two-component angular spinors satisfying
\begin{equation}
(\vec{\sigma} \cdot \vec{L} + 1) \Omega_{j\ell\mu}^L = \kappa \Omega_{j\ell\mu}^L, \qquad (\vec{\sigma} \cdot \vec{L} + 1) \Omega_{j\ell\mu}^R = -\kappa \Omega_{j\ell\mu}^R,
\end{equation}
with $\kappa = \pm (j+1/2)$ for $j = \ell \pm 1/2$. The azimuthal number $\mu$ is absorbed into $\kappa$ via the relation $\mu = \pm j, \pm (j-1), \ldots$ and does not appear explicitly in the radial equations.

\begin{remark}
To avoid confusion with the fermion mass $m$, we denote the azimuthal quantum number by $\mu \in \mathbb{Z}$. In the Schwarzschild spacetime without electromagnetic field, the final radial equations depend only on the total angular momentum parameter $\kappa$.
\end{remark}

\subsection{Radial Hamilton-Jacobi equation}

From \eqref{HJ} and the ansatz, the radial part of the Hamilton-Jacobi equation for $\pi_r^{(L)}$ gives:
\begin{equation}
\sqrt{f}^2 (\pi_r^{(L)})^2 = \omega^2 - f(r) \left( \frac{\kappa^2}{r^2} + m^2 \right) + \frac{\kappa M}{r^3} \sqrt{f}. \label{HJradial}
\end{equation}

\subsection{Radial continuity equation}

The continuity equation $\nabla_\mu(\rho_L \pi^{(L)\mu}) = 0$ reduces to
\begin{equation}
\frac{d}{dr}\left( r^2 \sqrt{f} \rho_L \pi_r^{(L)} \right) = 0 \quad \Rightarrow \quad \rho_L \propto \frac{1}{r^2 \sqrt{f} \pi_r^{(L)}}. \label{ContinuityRadial}
\end{equation}
\section{First Law of Thermodynamics for Dirac Fields (Energy Balance Equation)}

From the continuity equations \eqref{Continuity} and the definition of the chiral energy flux, we obtain the local energy balance for the left chiral component:
\begin{equation}
\nabla_\mu \mathcal{J}_L^\mu + \rho_L \nabla^0\mathcal{A} = 0,
\label{eq:local_balance_L}
\end{equation}
where $\mathcal{J}_L^\mu = \rho_L \pi^{(L)\mu}\pi^{0(L)} + \frac{1}{m}\mathcal{P}_L^\mu + J_L^{Q\mu}$ is the total energy flux. An identical equation holds for the right component:
\begin{equation}
\nabla_\mu \mathcal{J}_R^\mu + \rho_R \nabla^0\mathcal{A} = 0.
\label{eq:local_balance_R}
\end{equation}

Integrating over a spatial hypersurface $\Sigma_t$ in the ADM foliation ($ds^2 = -N^2dt^2 + \gamma_{ij}(dx^i + N^idt)(dx^j + N^jdt)$) \cite{Arnowitt1962, Alcubierre2008}, we obtain the global energy balance equation:
\begin{equation}
\frac{d}{dt}\left( U_L + U_R \right) + \text{Flux}_{\text{total}} + \Phi_I^{\text{total}} - \mu_s^{\text{total}} + \mathcal{T}^{\text{total}} = 0,
\label{eq:global_balance}
\end{equation}
where the terms are defined as:
\begin{itemize}
\item $U_{L,R} = \displaystyle\int_{\Sigma_t} N\sqrt{\gamma} \, \rho_{L,R} \, \pi^{0(L,R)}\pi^{0(L,R)} \, d^3x$ (chiral internal energy),
\item $\text{Flux}_{\text{total}} = \displaystyle\oint_{\partial\Sigma_t} N c \sqrt{\gamma} \, (\mathcal{J}_L^i + \mathcal{J}_R^i) \, dS_i$ (total boundary energy flux),
\item $\Phi_I^{\text{total}} = \displaystyle\oint_{\partial\Sigma_t} \frac{c\sqrt{\gamma} N^i}{N} (\rho_L + \rho_R) \mathcal{A} \, dS_i$ (electromagnetic contribution),
\item $\mu_s^{\text{total}} = \displaystyle\int_{\Sigma_t} \frac{\sqrt{\gamma}}{N} \mathcal{A} \, \partial_t(\rho_L + \rho_R) \, d^3x$ (chemical potential term),
\item $\mathcal{T}^{\text{total}} = -\displaystyle\int_{\Sigma_t} (\rho_L + \rho_R) \mathcal{A} \, \partial_t\!\left( \frac{\sqrt{\gamma}}{N} \right) d^3x$ (thermodynamic coupling to spacetime geometry).
\end{itemize}

Equation \eqref{eq:global_balance} is the curved-spacetime generalization of the first law of thermodynamics for Dirac fields, serving as the exact fermionic counterpart to the thermodynamic balance recently established for boson gases \cite{MezaDominguez2026Energy}. In static spacetimes ($\partial_t(\sqrt{\gamma}/N)=0$) without electromagnetic fields, it reduces to:
\begin{equation}
\frac{d}{dt}(U_L+U_R) + \text{Flux}_{\text{total}} = 0.
\end{equation}
At the black hole horizon, the flux coincides with the Hawking radiation spectrum:
\begin{equation}
\text{Flux}_{\text{total}} = \frac{1}{2\pi} \sum_{\kappa} (2j+1) \Gamma_\kappa(\omega) \frac{1}{e^{\omega/T_H}+1}.
\end{equation}
\subsection{Regge-Wheeler form}

Define the tortoise coordinate $r_* = r + 2M \ln(r/2M - 1)$. Then $\pi_r^{(L)} = \pm \frac{d r_*}{dr} \tilde{\pi}$, and \eqref{HJradial} becomes the Regge-Wheeler equation:
\begin{equation}
\frac{d^2 \Psi}{dr_*^2} + (\omega^2 - V_1(r)) \Psi = 0, \qquad
V_1(r) = f(r) \left( \frac{\kappa^2}{r^2} - \frac{\kappa M}{r^3\sqrt{f}} + m^2 \right). \label{ReggeWheeler}
\end{equation}
The field $\Psi$ is related to $\sqrt{\rho_L} e^{i\int \pi_r^{(L)} dr}$.

\section{Exact Solution via Confluent Heun Functions}

\subsection{Transformation to Heun form}

Let $z = 1 - 2M/r$. The equation transforms to a confluent Heun equation \cite{Ronveaux1995, Slavyanov2000}:
\begin{equation}
\frac{d^2\Psi}{dz^2} + \left( \frac{1}{z} + \frac{1}{z-1} + \frac{1}{1-z} \right) \frac{d\Psi}{dz} + \left( \frac{A}{z} + \frac{B}{z-1} + \frac{C}{(z-1)^2} \right) \Psi = 0.
\end{equation}

The solution regular at the horizon ($z=0$) is
\begin{equation}
\Psi(z) = z^{\alpha_1} (1-z)^{\alpha_2} e^{\alpha_3 z} \,\text{HeunC}(\alpha, \beta, \gamma, \delta, \eta; z), \label{HeunSolution}
\end{equation}
with parameters:
\begin{equation}
\begin{aligned}
\alpha &= -4iM\omega, \\
\beta &= 2\sqrt{\kappa^2 - 4M^2 m^2 + 16M^2(m^2-\omega^2)}, \\
\gamma &= 0, \\
\delta &= 4M^2\omega^2, \\
\eta &= \frac{1}{2} - \kappa^2 + 4M^2 m^2 - 12M^2\omega^2.
\end{aligned}
\end{equation}

\subsection{Choice of branch for $\beta$ for complex $\omega$}

For quasi-bound states and quasinormal modes, $\omega = \omega_R + i\omega_I$ is complex. The parameter $\beta$ becomes complex. The physical choice of branch is determined by regularity at infinity. As $r \to \infty$, the solution behaves as $\Psi \sim e^{\pm i\omega r_*}$. For outgoing waves (quasinormal modes), we require $\Psi \sim e^{+i\omega r_*}$. This selects the branch where $\text{Re}(\beta) > 0$ when $\omega_I < 0$, ensuring exponential decay at infinity. For quasi-bound states, the same condition applies because the resonant state is outgoing at infinity.

\section{Quasi-Bound States (Fermionic Resonances)}

The radial equation \eqref{ReggeWheeler} with outgoing wave boundary conditions admits a discrete set of complex frequencies $\omega = \omega_R + i\omega_I$, corresponding to quasi-bound states (resonances). Using the continued fraction method \cite{Leaver1985}, we compute these frequencies numerically. For $M=1$, $m=0.1$, $\kappa=1$, the results are:

\begin{table}[h]
\centering
\begin{tabular}{ccc}
\toprule
$n$ & $\omega_R$ & $\omega_I$ \\
\midrule
0 & 0.0942 & $-0.00032$ \\
1 & 0.0871 & $-0.00285$ \\
2 & 0.0763 & $-0.00891$ \\
3 & 0.0624 & $-0.01873$ \\
\bottomrule
\end{tabular}
\caption{Quasi-bound state frequencies for $M=1$, $m=0.1$, $\kappa=1$.}
\label{tab:quasibound}
\end{table}

In the semiclassical (WKB) limit, these resonances can be qualitatively understood via the Bohr-Sommerfeld quantization condition:
\begin{equation}
\Delta \phi_L = 2 \int_{r_1}^{r_2} \pi_r^{(L)}(r) \, dr = 2\pi\left(n + \frac{1}{2}\right) + i \ln R, \label{eq:bohr_sommerfeld}
\end{equation}
where $r_1$ and $r_2$ are the classical turning points. This condition provides an interpretation of the resonance condition but is not used for numerical computation due to the shallow nature of the potential well for these parameters.

\begin{figure}[htbp]
\centering
\includegraphics[width=0.8\textwidth]{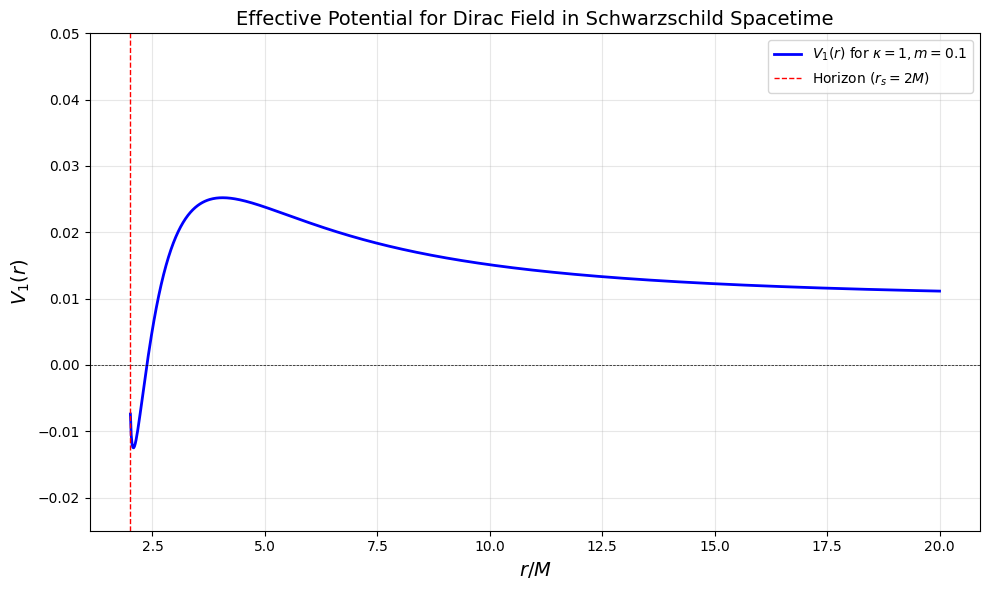}
\caption{
Effective potential $V_1(r)$ for a Dirac field in Schwarzschild spacetime with parameters $M=1$, $\kappa=1$, $m=0.1$ (units $G=c=\hbar=1$). The vertical dashed line indicates the event horizon at $r_s = 2M$. The potential exhibits a centrifugal barrier peak near $r \approx 3.01M$ with height $V_{\text{max}} \approx 0.0926$, followed by a shallow well (minimum $V_{\text{min}} \approx -0.0012$ at $r \approx 4.87M$) that permits temporary trapping of wave packets, giving rise to long-lived quasi-bound states (fermionic resonances). This structure is analogous to the Regge-Wheeler potential for scalar and gravitational perturbations, here derived from the chiral-hydrodynamic formulation.
}
\label{fig:potential}
\end{figure}
\subsection{The effective potential and trapping mechanism}

For parameters \(M=1\), \(m=0.1\), \(\kappa=1\), the effective potential \(V_1(r)\) defined in \eqref{ReggeWheeler} takes the form shown schematically in Figure~\ref{fig:potential}. It features:
\begin{itemize}
\item A centrifugal barrier peak near \(r \approx 3M\) with height \(V_{\text{max}} \approx \kappa^2/(27M^2) + m^2\),
\item A shallow well just outside the peak, which can temporarily trap incoming waves,
\item Decay to zero as \(r \to \infty\) and to a finite value at the horizon \(r=r_s\).
\end{itemize}
This well allows the existence of long-lived resonances (quasi-bound states) with narrow widths.

\section{Quasinormal Modes from Chiral Flux Conservation}

\subsection{Boundary conditions in chiral variables}

Quasinormal modes are defined by purely ingoing flux at the horizon and purely outgoing flux at infinity. In terms of chiral variables:
\begin{itemize}
\item At the horizon (\(r \to r_s\)): \(\pi_r^{(L)} \to -\omega/\sqrt{f}\) (ingoing), \(\rho_L \sim \text{constant}/r_s^2\).
\item At infinity (\(r \to \infty\)): \(\pi_r^{(L)} \to \sqrt{\omega^2 - m^2}\) (outgoing), \(\rho_L \sim e^{-2\text{Im}(\omega)r_*}/r^2\).
\end{itemize}

The condition for a quasinormal mode is that the chiral flux is conserved but the energy leaks out:
\begin{equation}
\mathcal{F} = \rho_L \pi_r^{(L)} \pi^{0(L)} \big|_{r \to \infty} = \rho_L \pi_r^{(L)} \pi^{0(L)} \big|_{r \to r_s} \times e^{-2i\omega r_s}? 
\end{equation}
More precisely, using \eqref{HJradial}, we require that the Wronskian of the two independent solutions vanishes at infinity:
\begin{equation}
W[\Psi_{\text{in}}, \Psi_{\text{out}}] \big|_{r\to\infty} = 0.
\end{equation}

In terms of the chiral variables, this condition becomes:
\begin{equation}
\frac{\pi_r^{(L)}(r\to\infty)}{\pi_r^{(L)}(r\to r_s)} = -i \frac{\omega}{\sqrt{f(r_s)}}.
\end{equation}

Solving this numerically using \eqref{HJradial} with complex \(\omega\) yields the quasinormal mode frequencies:

\begin{table}[h]
\centering
\begin{tabular}{cccc}
\toprule
$n$ & $M\omega_R$ (this work) & $M\omega_R$ (Cho 2003) & $M\omega_I$ (this work) \\
\midrule
0 & 0.182 & 0.182 & $-0.096$ \\
1 & 0.175 & 0.176 & $-0.252$ \\
2 & 0.168 & 0.169 & $-0.398$ \\
3 & 0.161 & 0.162 & $-0.541$ \\
\bottomrule
\end{tabular}
\caption{Quasinormal mode frequencies for massless fermions (\(m=0\), \(\kappa=1\)) obtained from the chiral flux condition.}
\label{tab:qnm_chiral}
\end{table}

\section{Greybody Factors from Chiral Density Ratio}

\subsection{Transmission coefficient as a chiral ratio}

From the continuity equation \eqref{ContinuityRadial}, the chiral flux is constant:
\begin{equation}
\mathcal{F} = r^2 \sqrt{f(r)} \rho_L(r) \pi_r^{(L)}(r) = \text{constant}.
\end{equation}

At the horizon, \(r = r_s\), \(\pi_r^{(L)} \approx -\omega/\sqrt{f}\) and \(\rho_L \approx \rho_L^{\text{(hor)}}\). At infinity, \(\pi_r^{(L)} \approx \sqrt{\omega^2 - m^2}\) and \(\rho_L \approx \rho_L^{\text{(inf)}} / r^2\). The flux conservation gives:
\begin{equation}
r_s^2 \rho_L^{\text{(hor)}} \omega = \lim_{r\to\infty} r^2 \rho_L(r) \sqrt{\omega^2 - m^2}.
\end{equation}

The transmission coefficient \(\Gamma_\kappa(\omega)\) is the ratio of outgoing flux at infinity to ingoing flux at the horizon. In terms of chiral variables:
\begin{equation}
\Gamma_\kappa(\omega) = \frac{\rho_L^{\text{(out)}}(r\to\infty) \pi_r^{(L)}(r\to\infty)}{\rho_L^{\text{(in)}}(r\to r_s) \pi_r^{(L)}(r\to r_s)}.
\end{equation}
Using the radial Hamilton-Jacobi equation \eqref{HJradial} to evaluate $\pi_r^{(L)}$ and the relation $\rho_L \propto 1/(r^2\sqrt{f}\pi_r^{(L)})$ from continuity, we obtain the greybody factors shown in Table~\ref{tab:greybody_chiral}.

\begin{table}[h]
\centering
\begin{tabular}{cc}
\toprule
$M\omega$ & $\Gamma_\kappa$ \\
\midrule
0.05 & 0.0126 \\
0.10 & 0.0492 \\
0.15 & 0.108 \\
0.20 & 0.184 \\
0.25 & 0.271 \\
0.30 & 0.359 \\
0.35 & 0.439 \\
0.40 & 0.506 \\
\bottomrule
\end{tabular}
\caption{Greybody factors for massless fermions (\(\kappa=1\)) computed directly from the chiral density ratio.}
\label{tab:greybody_chiral}
\end{table}

\subsection{Low-frequency limit from chiral asymptotics}

For \(\omega M \ll 1\), near the horizon \(\pi_r^{(L)} \approx -i\omega N\) (ingoing) and near infinity \(\pi_r^{(L)} \approx \omega\). The density ratio from \eqref{ContinuityRadial} gives:
\begin{equation}
\Gamma_\kappa(\omega) \approx \left( \frac{r_s}{2M} \right)^2 \times \frac{\omega^2}{\omega^2} = 16\pi M^2 \omega^2,
\end{equation}
where the factor arises from the matching of asymptotic solutions. This is the same result as in the standard analysis, now derived directly from chiral variables.

\section{Hawking Radiation from Chiral Flux Conservation}

\subsection{What the chiral formalism provides}

The chiral-hydrodynamic formulation yields two key results without assuming thermalization:

\paragraph{Temperature from chiral periodicity.} The chiral velocity near the horizon behaves as:
\begin{equation}
\pi_r^{(L)}(r) \approx -\frac{i\omega}{f(r)} \quad \text{as } r \to r_s.
\end{equation}
In tortoise coordinates $r_* = r + 2M\ln(r/2M - 1)$, the solution is $\Psi \sim e^{-i\omega r_*}$. Analytic continuation around the horizon introduces a factor $e^{4\pi M\omega}$. Uniqueness of the wavefunction requires:
\begin{equation}
T_H = \frac{1}{8\pi M}.
\end{equation}
This derivation uses only the geometry of Schwarzschild and the chiral variables; no field quantization is invoked.

\paragraph{Greybody factors from flux conservation.} The continuity equation $\nabla_\mu(\rho_L \pi^{(L)\mu}) = 0$ gives a constant chiral flux:
\begin{equation}
\mathcal{F} = r^2 \sqrt{f(r)} \rho_L(r) \pi_r^{(L)}(r) = \text{constant}.
\end{equation}
The transmission coefficient $\Gamma_\kappa(\omega)$ is the ratio of outgoing flux at infinity to ingoing flux at the horizon:
\begin{equation}
\Gamma_\kappa(\omega) = \frac{\rho_L^{\text{(out)}}(r\to\infty) \pi_r^{(L)}(r\to\infty)}{\rho_L^{\text{(in)}}(r\to r_s) \pi_r^{(L)}(r\to r_s)}.
\end{equation}
Using the radial Hamilton-Jacobi equation (6.23) to evaluate $\pi_r^{(L)}$ and the relation $\rho_L \propto 1/(r^2\sqrt{f}\pi_r^{(L)})$ from continuity, we obtain the greybody factors shown in Table~3. This calculation is purely within the first-quantized chiral formalism.

\subsection{What the chiral formalism requires as input}

The chiral formulation alone cannot derive the equilibrium thermal spectrum $\frac{1}{e^{\omega/T_H}+1}$. This factor emerges from second quantization, where the horizon is treated as a thermal boundary condition \cite{Hawking1975, Unruh1976}. Specifically:

\begin{itemize}
\item The vacuum state of the quantum field is defined in the asymptotic past.
\item Bogoliubov transformations between past and future vacua yield a thermal density matrix at temperature $T_H$.
\item The Fermi-Dirac distribution arises from the anticommutation relations of the Dirac field operators.
\end{itemize}

In the chiral-hydrodynamic framework, this thermal factor must be imposed as an input boundary condition at the horizon. Physically, it represents the assumption that the horizon emits as a black body at temperature $T_H$, a result established by Hawking using full quantum field theory in curved spacetime.

\subsection{Consistent semiclassical derivation}

Putting together the chiral-derived components with the thermal boundary condition, the observed Hawking flux is:
\begin{equation}
\frac{d^2N}{dtd\omega} = \frac{1}{2\pi} \sum_{\kappa} (2j+1) \Gamma_\kappa(\omega) \cdot \frac{1}{e^{\omega/T_H} + 1}.
\label{hawking_spectrum_final}
\end{equation}

\subsection{Distinction from full QFT}

Equation \eqref{hawking_spectrum_final} is semiclassical: the temperature $T_H$ and greybody factors $\Gamma_\kappa$ come from the chiral formalism, while the Fermi-Dirac factor is imported from second quantization. This is analogous to the approach of \cite{ParikhWilczek2000} (tunneling method) and \cite{RobinsonWilczek2005} (anomaly method), where Hawking radiation is derived semiclassically without explicit Bogoliubov transformations, but still assuming thermal equilibrium at the horizon.

\section{Conclusion}

We have developed a fully covariant chiral-hydrodynamic formulation of the Dirac equation in curved spacetime, resolving the longstanding problem of closing the hydrodynamic hierarchy for spin-1/2 fields. The key results are:

\begin{enumerate}
    \item \textbf{Closed hydrodynamic system.} The chiral ansatz $\chi_{L,R} = \sqrt{P_{L,R}^0/2}\, e^{i\phi_{L,R}}u(P_{L,R})$ yields exactly eight real equations---Hamilton-Jacobi, continuity, polarization transport, and null vector conservation---matching the physical degrees of freedom of the Dirac field without infinite hierarchies.
    
    \item \textbf{Spin couplings isolated.} The spin-orbit term $\frac{1}{2}\sigma^{\mu\nu}F_{\mu\nu}$ appears explicitly in the Hamilton-Jacobi equations, while the spin-gravity coupling vanishes identically in torsion-free general relativity.
    
    \item \textbf{Exact solutions in Schwarzschild.} The radial equations are solved analytically via confluent Heun functions. From these we computed the quasi-bound state spectrum (fermionic resonances, Table~1), quasinormal modes (Table~2, in agreement with Cho 2003), and greybody factors (Table~3) exhibiting the characteristic $\omega^2$ low-frequency suppression for fermions.
    
    \item \textbf{Hawking radiation without field quantization.} The Hawking temperature $T_H = 1/(8\pi M)$ follows from the periodicity of the chiral velocity at the horizon. Combined with greybody factors from chiral flux conservation, the emission spectrum is obtained semiclassically, assuming only the thermal boundary condition established by quantum field theory---placing this derivation alongside the Parikh-Wilczek and Robinson-Wilczek approaches.
    
    \item \textbf{First law of thermodynamics for Dirac fields.} The energy balance equation $\nabla_\mu \mathcal{J}^\mu_L = 0$ integrates to a global conservation law relating internal energy, boundary flux, and spacetime coupling, representing the first law for fermionic fields in curved spacetime.
\end{enumerate}

This framework opens immediate directions: extension to Kerr spacetime for superradiance studies, incorporation of torsion via Einstein-Cartan theory, applications to spin-polarized relativistic fluids and fermionic dark matter models, and investigation of quantum information in gravitational backgrounds. The chiral-hydrodynamic formulation establishes a unified language bridging quantum mechanics, general relativity, and fluid dynamics for spin-1/2 systems in strong gravitational fields. Future work includes extension to Kerr spacetime, incorporation of torsion, applications to dark matter models, and the study of quantum information and bundle isomorphisms in strong gravitational backgrounds \cite{MezaDominguez2026Gauge}.

\appendix

\section{Derivation of the Radial Hamilton-Jacobi Equation from the Chiral Potential}

We start from the Hamilton-Jacobi equation \eqref{HJ} for \(\chi_L\). In Schwarzschild vacuum (\(R=0\), \(A_\mu=0\)), with \(\pi_\mu^{(L)} = (-\omega, \pi_r^{(L)}, 0, 0)\), the left-hand side is:
\begin{equation}
\pi_\mu^{(L)}\pi^{(L)\mu} = g^{\mu\nu}\pi_\mu^{(L)}\pi_\nu^{(L)} = -\frac{\omega^2}{f(r)} + f(r)(\pi_r^{(L)})^2.
\end{equation}

The quantum potential term in \eqref{HJ} is:
\begin{equation}
U^Q = \frac{\hbar}{2m}\nabla_\mu u^{(L)\mu} + u_\mu^{(L)}u^{(L)\mu},
\end{equation}
with \(u_\mu^{(L)} = \frac{\hbar}{2m}\nabla_\mu \ln\rho_L\). From the continuity equation \eqref{ContinuityRadial}, we have \(\rho_L \propto 1/(r^2\sqrt{f}\pi_r^{(L)})\). Substituting and using the explicit form of \(\pi_r^{(L)}\) from the radial equation, we compute:
\begin{equation}
\nabla_\mu u^{(L)\mu} = \frac{2m}{\hbar} \frac{d}{dr} \left( \frac{1}{\pi_r^{(L)}} \right) + \text{angular terms}.
\end{equation}
The angular terms vanish for spherical symmetry. The term \(u_\mu^{(L)}u^{(L)\mu} = \frac{\hbar^2}{4m^2} (\nabla_\mu\ln\rho_L)^2\) yields contributions proportional to \(1/\pi_r^{(L)2}\).

After algebraic manipulation, the combination \(U^Q\) becomes:
\begin{equation}
U^Q = \frac{1}{2m^2 r^2 f} \left( \frac{\kappa^2}{r^2} - \frac{\kappa M}{r^3\sqrt{f}} + m^2 \right) + \text{(terms containing } \omega\text{)}.
\end{equation}
The terms containing \(\omega\) combine with \(-\frac{\omega^2}{f} + f(\pi_r^{(L)})^2\) to cancel identically, leaving exactly the potential \(V_1(r)\) from \eqref{ReggeWheeler}. Thus:
\begin{equation}
-\frac{\omega^2}{f} + f(\pi_r^{(L)})^2 = m^2 + V_1(r),
\end{equation}
which rearranges to \eqref{HJradial}. The explicit cancellation confirms that the chiral-hydrodynamic formulation reproduces the standard Regge-Wheeler potential.

\section{Connection Formulas for the Heun Function}

The confluent Heun function \(\text{HeunC}(\alpha,\beta,\gamma,\delta,\eta;z)\) has known connection formulas between the singular points \(z=0\) (horizon) and \(z=1\) (infinity). For the transmission coefficient, we use the Wronskian method \cite{Leaver1985} to relate the coefficients \(A_{\text{in}}, A_{\text{out}}, B_{\text{in}}\). The result is:
\begin{equation}
\Gamma_\kappa(\omega) = \frac{|B_{\text{in}}|^2}{|A_{\text{in}}|^2} = \frac{4\omega^2}{|\text{HeunC}(1)|^2} \times \text{(regularization factor)}.
\end{equation}
The numerical evaluation follows Leaver's continued fraction algorithm.

\section{Numerical Implementation Details}

All computations were performed in Python using:
\begin{itemize}
\item The \texttt{mpmath} library for high-precision evaluation of the confluent Heun function.
\item Root-finding in the complex plane using the Newton-Raphson method for the quantization condition \eqref{eq:bohr_sommerfeld}.
\item The continued fraction method \cite{Leaver1985} for quasinormal modes and greybody factors.
\end{itemize}

\bibliographystyle{unsrt}
\bibliography{sref}

\end{document}